\documentclass{article}
\usepackage{spconf,amsmath,graphicx,multirow}

\title{AN IMPROVED HYBRID CTC-ATTENTION MODEL FOR SPEECH RECOGNITION}
\name{Zhe Yuan, Zhuoran Lyu, Jiwei Li and Xi Zhou\sthanks{\{yuanzhe, lvzhuoran,  lijiwei, zhouxi\}@cloudwalk.cn}}
\address{Author Affiliation(s)}

\address{Cloudwalk Technology Inc, Shanghai, China}
\begin{document}
%
\maketitle
\begin{abstract}
Recently, end-to-end speech recognition with a hybrid model consisting of the connectionist temporal classification(CTC) and the attention encoder-decoder achieved state-of-the-art results. In this paper, we propose a novel CTC decoder structure based on the experiments we conducted and explore the relation between decoding performance and the depth of encoder. We also apply attention smoothing mechanism to acquire more context information for subword-based decoding. Taken together, these strategies allow us to achieve a word error rate(WER) of 4.43\% without LM and 3.34\% with RNN-LM on the test-clean subset of the LibriSpeech corpora, which by far are the best reported WERs for end-to-end ASR systems on this dataset.

\end{abstract}
\section{Introduction and Background}
\label{IB}

Automatic speech recognition (ASR), the technology that enables the recognition and translation of spoken language into text by computers, has been widely used  in different applications. In the past few decades, ASR relied on complicated traditional techniques including Hidden Markov Models (HMMs) and Gaussian Mixture Models (GMMs) \cite{b1}. Besides, these traditional models also require hand-made pronunciation dictionaries and predefined alignments between audio and phoneme\cite{a1,a2}. Although these traditional models achieve state-of-the-art accuracies on most audio corpora, it is quite a challenge to develop ASR models without enough acoustics knowledge. Therefore, benefiting from rapid development of deep learning, a few end-to-end ASR models were raised in recent years.

Connectionist temporal classification(CTC) based models and sequence-to-sequence(seq2seq) with attention models are two major approaches in end-to-end ASR systems. Both methods address the problem of variable-length input audios and output texts. Deep Speech 2, which was came up with by Baidu Silicon Valley AI Lab in 2016 \cite{b2}, making full use of CTC and RNN, achieved a state-of-the-art recognition accuracy. As for seq2seq model, Chorowski et al utilized seq2seq model with attention mechanism to perform speech recognition \cite{b3}. However, the accuracy of the model is unsatisfactory since alignment estimation in the attention mechanism is easily corrupted by noise, especially in real environment tasks. 

To overcome the above misalignment problem, a combination of CTC and attention-based seq2seq model were proposed by Watanabe in 2017 \cite{b4}. The key to this joint CTC-attention model is training a shared encoder, with both CTC and attention decoder as objective functions simultaneously. This novel approach improves the performance in both training speed and recognition accuracy. 

This paper is partly inspired by the above method. Our main contributions in this paper include exploring different encoder and decoder network architecture and adopting several optimization methods such as attention smoothing and L2 regularization. We demonstrate that our system outperforms other published end-to-end ASR models in WER on LibriSpeech dataset. 

The paper is organized as follows. Section \ref{RW} briefly introduces the related works, mainly focusing on the hybrid CTC/Attention method. Section \ref{M} details our model architecture and section \ref{E} presents our training methods and experimental results. Finally, section  \ref{C} concludes this work.

\section{Related Work}\label{RW}

In this section, we review the hybrid CTC-attention architecture in Section \ref{hca} and unit selection methods in Section \ref{bs}.

\subsection{Hybrid CTC-attention architecture}\label{hca}

\begin{figure}[htbp]
\centerline{\includegraphics[width=0.45\textwidth]{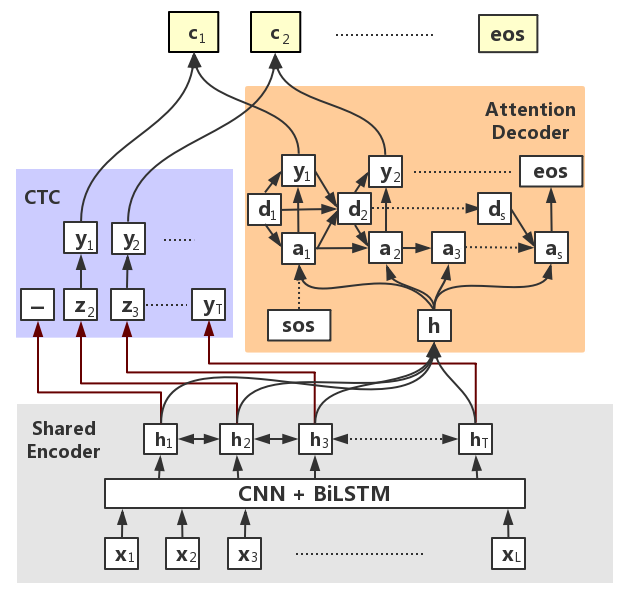}}
\caption{Architecture of the hybrid CTC-Attention model}
\label{f1}
\end{figure}

The idea of this architecture is to use CTC as an auxiliary objective function to train the attention-based seq2seq network. 
Fig.~\ref{f1} illustrates the architecture of the network, where the encoder has several convolutional neural network(CNN) layers followed by bidirectional long short-term memory (BiLSTM) layers, while the decoder includes a CTC module and an attention-based network. According to \cite{b5}, using CTC along with attention decoder brings more robustness to the network since CTC helps acquiring appropriate alignments in noisy conditions. Moreover, CTC also assists the network in training speed. 

CTC, which is introduced by \cite{b6}, provides a method to train RNNs without any prior alignments between inputs and outputs. 
Suppose the length of the input sequence is $t$, then the probability of a CTC path can be computed as follow:
\begin{equation}
p_{ctc}(c|x) = \prod_{t=1}^Tq_{t}^{c_t}\label{eq1}
\end{equation}
where $q_{t}^{c_t}$ denotes the the softmax probability of outputting label $c_t$ at frame t and $c=(c_1, c_2,..., c_t)$ denotes the CTC path. Hence the likelihood of the label sequence can be computed as follow:
\begin{equation}
p_{ctc}(y|x) = \sum_{c\in\theta(y)}p_{ctc}(c|x)
\end{equation}
where $\theta(y)$ is the set of all possible CTC paths that can be mapped to $y$. Therefore, we have CTC loss to be:

\begin{equation}
L_{ctc} = -\textit{ln}(p_{ctc}(y|x))\label{eq2}.
\end{equation}

As for decoder part, the possibility of label $y_s$ at each step depends on input feature $h$ and previous labels $y_{1:s-1}$. The overall possibility of the entire sequence can be obtained as follow:

\begin{equation}
p_{att}(y|x) = \prod_{s}p_{att}(y_s|h,y_{1:s-1})\label{eq3}
\end{equation}
where 
\begin{equation}
y_s = FullyConnected(d_s,a_s),
\end{equation}
\begin{equation}
d_s = LSTM(d_{s-1}, y_{s-1},a_s),
\end{equation}
\begin{equation}
a_s = \sum_t\epsilon_{s,t}h_t.
\end{equation}
$d_s$ denotes $LSTM$ hidden states while $a_s$ is the context vector based on input features $h$ and attention weight $\epsilon_{s,t}$ in the above equation. The loss function of this part is defined as:
\begin{equation}
L_{att} = \alpha L_{ctc} + (1-\alpha)L_{att}\label{eq8}
\end{equation}
where $\alpha$ denotes the weight of different loss, $\alpha \in[0,1]$.

\subsection{Unit selection}\label{bs}
Methods based on large lexicon, such as phoneme-based ASR systems or word-based ASR systems, are not able to resolve out-of-vocabulary (OOV) problems. Thus, starting from LAS \cite{b7}, such seq2seq model raises new character-based method. By combining frame information in audio clips and the corresponding characters together, the OOV problem is resolved to some extent. Since many characters in English words are silent and same characters in different sentences may pronounce differently (e.g. "a" in "apple" and "approve"), decoding procedure on character level relies heavily on the sentence sequence relationship given by RNN rather than the acoustic information given by the audio clip frames, which results in the uncertainty of decoding procedure on character level. Considering all the issues mentioned above, subword-based structure can resolve OOV problems on one hand, and can learn the relationship between acoustic information and character information on the other hand. An effective and fast method for generating subwords is byte-pair encoding (BPE) \cite{b8}. Which is a compression algorithm that iteratively replaces the most frequent pair of units (or bytes) with an unused unit, and eventually generates new units that are consistent with the number of iterations.

\section{Methodology}\label{M}

In this section, we detail our optimization and improvements based on the previous hybrid CTC-attention architecture. We show our improvements to encoder-decoder architecture and attention mechanism in section \ref{ea} and section \ref{sa}.

\subsection{Encoder-Decoder architecture}\label{ea}
\begin{figure}[htbp]
\centerline{\includegraphics[width=0.5\textwidth]{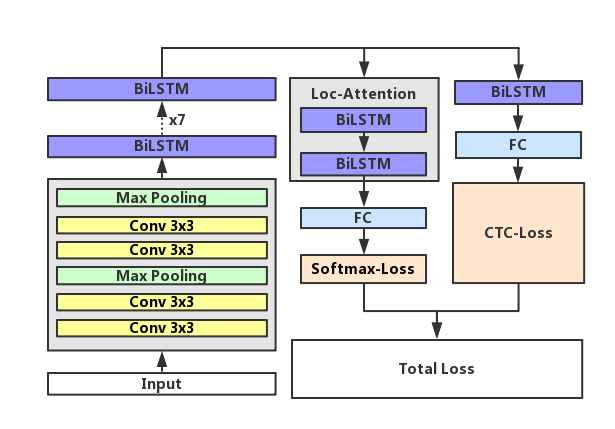}}
\caption{Encoder and decoder architecture of our model}
\label{f2}
\end{figure}

The authors in Espnet \cite{a3}, stacked several BiLSTM layers above a few convolutional layers. The outputs of the last BiLSTM layer sever as inputs to both CTC and attention-decoder as shown in Fig.~\ref{f1}. Our major improvements conclude inserting a BiLSTM layer, which is solely occupied by the CTC branch, between the top shared encoder layer and FC layer connected to CTC. The entire hybrid architecture is shown in Fig.~\ref{f2}.

According to our experiments in Section \ref{E}, setting $\alpha$ in \eqref{eq8} to a smaller value makes the network perform better. 
However, when the weight $\alpha$ is low, a new problem is raised. Since lower $\alpha$ brings smaller gradient descent in back propagation in CTC loss part, the shared decoder focuses more on the attention module than the CTC module, which limits the performance of the CTC decoder. Considering this limitation, we introduce a solely BiLSTM layer linking to the CTC decoder, which can compensate the problem we mentioned above.

\subsection{Attention smoothing}\label{sa}

Inspired by \cite{b3}, we use a location-based attention mechanism in our implementation. Specifically, the location based attention energies $e_{s,t}$ can be computed by the following equation:
\begin{equation}
e_{s,t}= W^Ttanh(W_dd_{s-1}+W_hh_t+W_ff_{s,t}+b)
\end{equation}
where 
\begin{equation}
\epsilon_{s,t}=exp(e_{s,t})/\sum_{i=1}^Texp(e_{s,i})
\end{equation}
and $f_s=F*\epsilon_{s-1}$.

In our speech recognition system, subwords are chosen as the model units, which require more sequence context information than character-based units. However, the attention score distribution is usually very sharp when computed using above equations. Hence, we apply attention smoothing mechanism instead, which can be computed by
\begin{equation}
\epsilon_{s,t}=sigmoid(e_{s,t})/\sum_{i=1}^{T}sigmoid(e_{s,i}).
\end{equation}
The above method successfully smooths attention score distributions and then keep more context information for subword-based decoding.

\section{Experiments}\label{E}
\subsection{Experimental Setup}\label{ES}
We train and test our implementation over LibriSpeech dataset \cite{b9}. Specificlly, we use train-clean-100, train-clean-360, train-other-500 as our training set and dev-clean as our validation set. For evaluation, we report the word error rates (WERs) on the subsets test-clean, test-other, dev-clean and dev-other. We also adopt 3-fold speed perturbation(0.9x/1.0x/1.1x) for data augmentation. 80 dimensional Mel-filterbank features are generated using a sliding window of length 25 ms with 10ms stride, and the feature extraction is performed by KALDI toolkit \cite{b10}. Subword units are extracted using all the transcripts of training data by BPE algorithm. The number of subword units is set to 5000.

We use a 4-layer CNN architecture followed by a 7-layer BiLSTM where each layer is a BiLSTM with 1024 cell units per direction as encoder. In the CNN part, input features are downsampled to 1/4 through two max-pooling layers. The decoder consists of two branches where one branch is a one-layer BiLSTM followed by a CTC decoder and the other branch is a 2-layer LSTM with 1024 cell units per layer.

The AdaDelta algorithm \cite{b11} with initial hyper-parameter epsilon=1e-8 is used for optimization, and L2 regularization and gradient clipping are applied. We measure the accuracy of the validation set every 1000 iterations and apply a strategy that eps is decayed by 0.1 when the average validation accuracy drops. All experiments are performed on 4 Tesla P40 GPUs with batchsize $=40$ on each GPU.

Our language model is a two-layer LSTM with units=1536 trained on large text data of 14500 public domain books, which is commonly used as training material for the LibriSpeech's LM. The SGD algorithm is used for optimization, with initial learning-rate 1.0 and lr-decay 0.9 per 2 epochs. For decoding, we use the beam search algorithm with the beam size 20.

\subsection{Results}\label{R}
\begin{figure}[htbp]
\centerline{\includegraphics[width=0.5\textwidth]{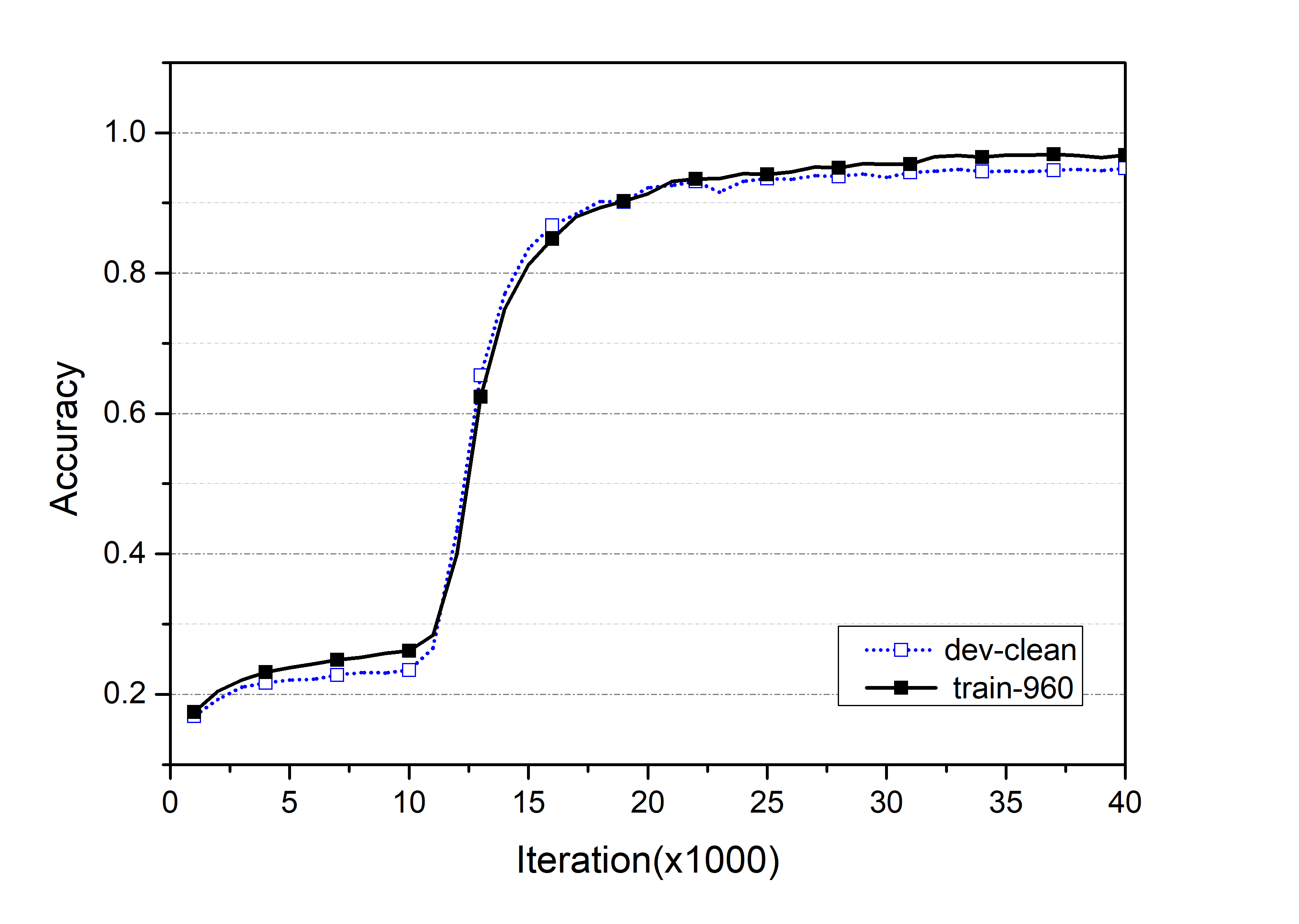}}
\caption{Accuracy curves with the number of iterations on both the train set and the validation set during training}
\label{f3}
\end{figure}

Fig.~\ref{f3} shows the accuracy curve during training process, from which we can see that the model converges after 35000 iterations. The perplexity of our trained RNN-LM is 50.4 on the training set, and 46.9 on the test-clean subset.  We conduct experiments with different number of layers in encoder and the addition of BiLSTM layer on the CTC branch. Results are shown in Table.~\ref{tab1}, from which we have the following comments. Both increasing the number of BiLSTM layers in encoder and the addition of BiLSTM layer on the CTC branch lead to better WER. Moreover, WER can be reduced by about 25\% using our trained RNN-LM.

\begin{figure}[htbp]
\centerline{\includegraphics[width=0.5\textwidth]{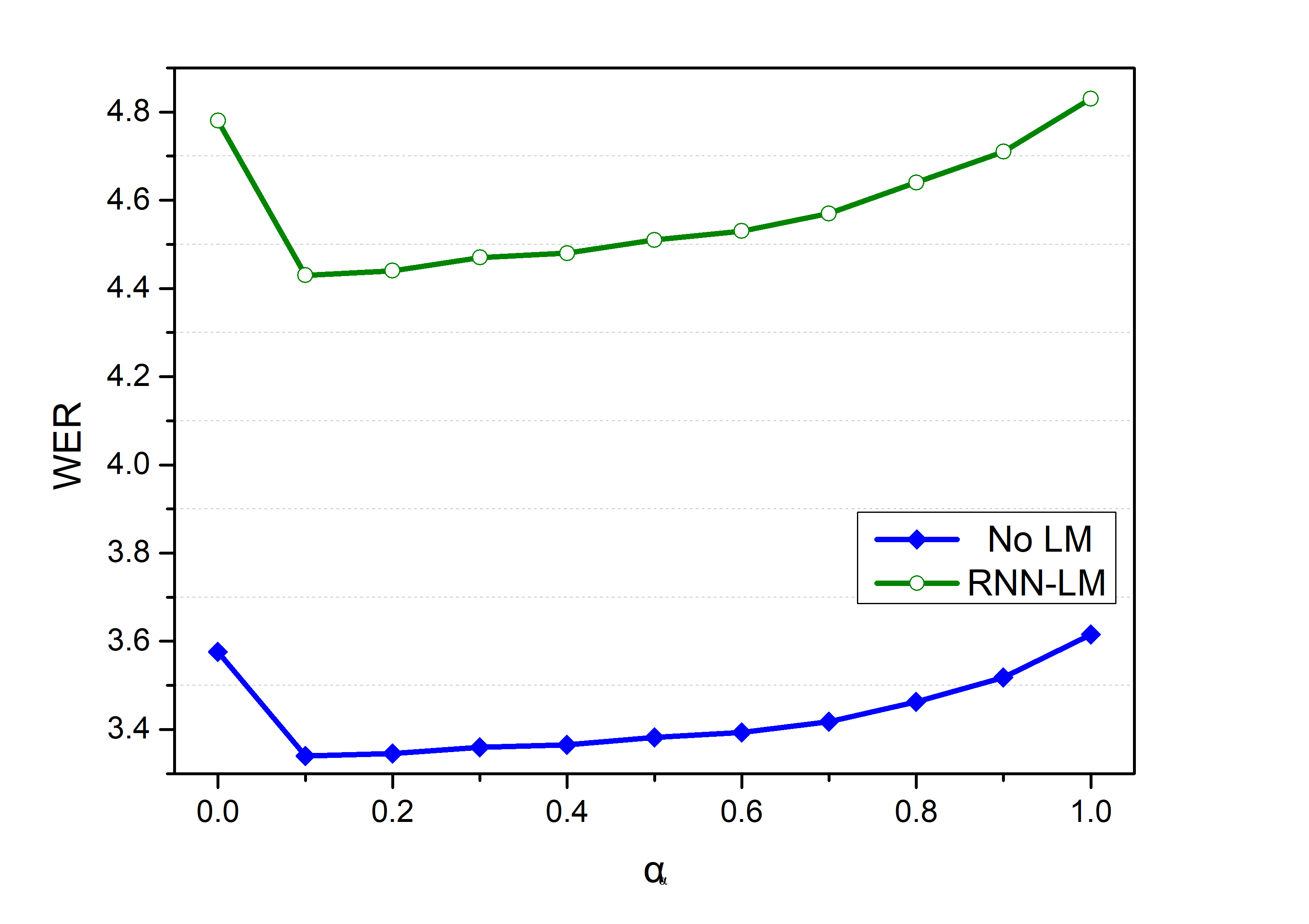}}
\caption{WER performance as a function of alpha on test-clean subset}
\label{f4}
\end{figure}

\begin{table}[htbp]
\caption{Comparison of test-clean-subset WERs under different structures}
\begin{center}
\begin{tabular}{ccc}
\hline
Encoder Layers& WER(no LM) & WER(RNN-LM) \\
\hline
5&  5.01 & 3.73\\
5+CTC-BiLSTM& 4.73 & 3.59 \\
6& 4.82& 3.64\\
6+CTC-BiLSTM& 4.57 & 3.43\\
7& 4.64 & 3.51\\
7+CTC-BiLSTM& 4.43 & 3.34 \\
\hline
\end{tabular}
\label{tab1}
\end{center}
\end{table}

\begin{table}[thbp]
\caption{Performance of different networks on the LibriSpeech dataset}
\begin{center}
\begin{tabular}{|c|c|c|c|c|}
\hline
\multirow{2}*{\textbf{Model}}& \multicolumn{2}{|c|}{\textbf{Test}}& \multicolumn{2}{|c|}{\textbf{Dev}} \\
\cline{2-5} 
&\textit{clean}& \textit{other}& \textit{clean}& \textit{other}\\
\hline
Baidu DS2\cite{b2} + LM& 5.15& 12.73& -& - \\
Espnet\cite{b12} + LM& 4.6& 13.7& 4.5& 13.0 \\
I-Attention\cite{b13} + LM & 3.82& 12.76& 3.54& 11.52 \\
\hline
Ours + no LM& 4.43& 13.5& 4.37& 13.1 \\
Ours + LM& \textbf{3.34}& \textbf{10.54}& \textbf{3.15}& \textbf{9.98} \\
\hline
\end{tabular}
\label{tab2}
\end{center}
\end{table}

After that, we compare the different weight $\alpha$ between CTC loss and attention loss by step 0.1. The result is shown in Fig.\ref{f4}. When we use pure attention-based system or pure ctc-based system, it produces inferior performance. The curve also shows that decreasing $\alpha$ leads to better WER in hybrid system, which is consistent with the purpose of using CTC-decoder at the beginning: the CTC module is mainly used to assist the monotonic alignment and increase the convergence speed of training, and the hybrid system decoding effect mainly relies on the attention-decoder. As fig.~\ref{f4} shows, we find that the best tuned $\alpha$ is 0.1. 
    
Finally, we compare our results with other reported state-of-the-art end-to-end systems on the LibriSpeech dataset in Table.~\ref{tab2}. The results show that our system achieves better WERs than other known end-to-end ASR models.


\section{Conclusions}
\label{C}

In summary, we explore a variety of structural improvements and optimization methods on the hybrid CTC-attention-based ASR system. By applying the CTC-decoder BiLSTM, attention smoothing and some other tricks, our system achieves a word error rate(WER) of 4.43\% without LM and 3.34\% with RNN-LM on the test-clean subset of LibriSpeech corpus.

Future work will concentrate on the optimization of both the decoder structure and the training method, such as finetuning the CTC-decoder-branch after training the shared encoder. Another future work is to apply this technique to other languages like Mandarin, in which there are many polyphonic words that need to be solved in the decoding process.

\bibliographystyle{IEEEbib}
\bibliography{strings,refs}

\end{document}